\documentstyle[11pt,newpasp,twoside]{article}
\begin{document}

\title{Summary: Modes of Star Formation}

\author{Richard B. Larson}

\affil{Yale Astronomy Department, New Haven, CT 06520-8101, USA}
\bigskip

\section{Introduction}

   This meeting has featured many interesting developments in a wide
range of topics related to star formation, and it will not be possible
for me to review all of them in this brief summary; instead, I shall
focus on what seemed to me some of the main themes or conclusions of the
meeting.  The title ``Modes of Star Formation'' can refer to a variety
of different aspects of star formation, and it can even mean different
things to different people; we heard, for example, about `isolated'
and `clustered' modes of star formation, `spontaneous' and `triggered'
modes, `quiescent' and `starburst' modes, and even about `low-mass' and
`high-mass' modes.  Perhaps the most general type of question addressed
by this subject is how star formation is organized, and what kinds of
patterns we can discern in how and where it occurs.

\section{Star-Forming Clouds and Cores}

   It has long been known that stars form in molecular clouds, and
several presentations at this meeting dealt with molecular clouds and
their structure and dynamics.  The gross properties of these clouds,
such as their sizes, masses, and linewidths, seem to have reasonably
well defined values and to follow similar trends in different regions,
and even in different galaxies.  Molecular clouds thus exhibit some
systematic properties that we might expect to be reflected in
corresponding systematic features of the way in which stars form.
But it is also clear on closer inspection that star formation must be
a complex and even somewhat chaotic process, since molecular clouds
are quite irregular in their structure and have supersonic internal
turbulent motions.  Progress in understanding the details of star
formation has therefore been slow, and we do not yet have a theory
with much predictive power.

   Much effort has gone into understanding the origin and decay of
the turbulence in molecular clouds, since this turbulence is the main
effect counteracting gravity and since star formation cannot occur until
it has been dissipated, at least locally.  It has often been assumed
that molecular clouds are long-lived quasi-equilibrium structures, and
the apparent persistence of turbulence in them for many crossing times
has been considered problematic.  Magnetic fields have been thought
to solve this problem by prolonging the dissipation time, but recent
numerical simulations have shown that turbulence always decays within
a crossing time, even with magnetic fields.  The properties of molecular
clouds are therefore hard to understand unless these clouds are
short-lived, and indeed it has become increasingly clear that
star-forming clouds are transient structures.  The ages of the young
stars and clusters associated with them imply that star formation is
a rapid process and continues for only about a crossing time in each
region, after which the remaining gas is quickly dispersed; for example,
clusters of stars older than 5~Myr have already cleared away their
surrounding gas, while clusters older than 10~Myr no longer have any
associated gas within many parsecs.  The weak-lined T~Tauri stars
discussed at this meeting provide further evidence for rapid cloud
dispersal, since these stars were evidently formed in situ from gas that
has already disappeared after only a few Myr.  If star-forming clouds
are transient, there is no longer any problem in understanding why they
are turbulent, since they must then be condensations in a turbulent
medium that are continually forming and dispersing and that don't last
long enough for their turbulence to be completely dissipated.

   Various scales of structure have been distinguished in molecular gas,
including features called `clouds', `clumps', and `cores', which form
stellar groupings of various sizes including associations, clusters,
and individual stars or binary systems.  Much study has been devoted
to the apparently most fundamental star-forming units, the cloud cores,
which are individual density peaks with relatively little internal
structure.  In nearby clouds, these cores have sizes of the order of
0.1~pc and masses of the order of one solar mass.  Some of them contain
embedded young stars, and others show evidence for infall in their
line profiles which suggests that they are currently collapsing. The
inferred infall motions, as we heard, are too large in both amplitude
and spatial extent to be consistent with standard cloud models that
assume slow quasi-static evolution, and instead they favor a more
dynamical picture in which cores form and collapse rapidly into stars.
Moreover, the mass spectrum of the cores in the well-studied $\rho$~Oph
cloud is quite similar to the stellar IMF, and this suggests that these
cores are the direct progenitors of stars and that the stellar IMF is
determined, at least in part, by the mass spectrum of the cloud cores.

\section{The Stellar IMF as a Fossil Record}

   Many presentations at this meeting dealt with the stellar Initial
Mass Function found in different regions and systems.  The apparent
universality of the IMF has been emphasized in recent years, and the
IMF does indeed show an impressive degree of uniformity in systems that
include the local field population, nearby star-forming regions, star
clusters in our Galaxy and the Magellanic Clouds, and nearby dwarf
galaxies.  The resulting `standard IMF' is now fairly well defined,
and it resembles the original Salpeter power law at masses above a
solar mass but flattens below a solar mass and then probably declines
(in logarithmic units) in the brown dwarf regime.  The departure of
the IMF from a power law below a solar mass is beyond question, and
it appears instead that the IMF is a broadly peaked function with a
characteristic mass of the order of one solar mass.  This characteristic
stellar mass is similar to the typical mass of the dense cores in nearby
clouds, and also to the Jeans (or Bonnor-Ebert) mass predicted for cloud
fragmentation at the temperatures and pressures typically observed in
star-forming clouds.  This suggests that the IMF is determined by cloud
fragmentation processes that have a mass scale similar to the Jeans
mass. 

   However, despite this evidence for uniformity, there is also evidence
for departures from a standard IMF that is becoming increasingly
difficult to ignore.  For example, the Taurus region has a deficiency of
brown dwarfs and possibly also of stars above a solar mass, suggesting
that star formation in simple environments like the Taurus clouds
produces only a limited range of masses and that stars with smaller and
larger masses form only under conditions that are not present in Taurus.
For example, brown dwarfs might form mostly by subfragmentation in more
complex environments, while massive stars might form mostly in larger
groupings or clusters.  In fact, it has become increasingly clear that
massive stars do form preferentially if not exclusively in clusters, and
that they are most likely to form at the centers of massive and dense
clusters like the Trapezium cluster.  We heard about several examples
of massive clusters with apparently top-heavy IMFs; for example, in the
30~Doradus cluster the IMF appears to flatten below about 2~M$_\odot$,
i.e.\ at a somewhat higher mass than the standard IMF, while in the
extremely luminous cluster M82-F the IMF may flatten below 3 or
4~M$_\odot$.  The massive young Arches and Quintuplet clusters recently
discovered near the Galactic center also have relatively flat IMFs, so
there may indeed be a tendency for massive clusters to form
preferentially massive stars.

\section{Formation of Massive Stars}

   The formation of massive stars has itself become a topic of major
interest, and several presentations discussed the relevant observational
evidence and theoretical ideas.  Massive stars appear to form in
exceptionally dense environments, and two competing hypotheses are that
they form by gas accretion and that they form by stellar coalescence.
Neither of these hypotheses can yet be excluded, and a further
intermediate possibility suggested here is that interactions between
dense star-forming cores are involved.  In fact, such a picture seems
almost unavoidable, because if one imagines that accretion is the
dominant process, then the gas being accreted by the stars in a forming
cluster must be very clumpy and must contain many forming stars, while
if one imagines that coalescence is involved, the coalescing stars will
still have massive gas envelopes that must also play a role.  Thus, one
is led in either case to a picture in which interactions between forming
stars or protostars in dense cluster-forming cloud regions are involved.
If the accumulation processes that build up massive stars are scale-free
and if no new scale larger than the Jeans mass enters the problem, then
a power-law upper IMF could plausibly result, but a quantitative theory
of such processes is still in the future.

   An extreme case of massive star formation may occur with the 
`Population~III' stars that form at very early times before any heavy
elements have been produced.  The gas temperature is then much higher
than in present-day molecular clouds, and the Jeans mass is therefore
also much higher.  The recent simulations of early star formation
presented here suggest that the first stars might have had masses
between 30 and 300 solar masses.  These objects would have had important
consequences for ionizing and chemically enriching the early universe.
Another intriguing possibility suggested here was that in the presence
of a very intense radiation field, such as exists at the center of M51,
the temperature and Jeans mass may again be very high, possibly allowing
the formation of isolated massive stars.  The central starburst region
of M82 has long been suspected to harbor a top-heavy IMF (some evidence
for which may have been found in the case of M82-F), and it will be
interesting to see if further studies of star formation in extreme
environments reveal further examples of anomalous and possibly
top-heavy IMFs.

\section{Binaries, Clusters, and Associations}

   The clustering of young stars on scales ranging from binaries to
large associations provides another record of the way in which stars
form.  Like stellar masses, the properties of binaries are largely
preserved from their time of formation, and this allows binary
statistics to be used to infer something about the typical sites of star
formation.  The field population contains a mix of contributions from
star-forming regions of all types, and since the frequency of binaries
is observed to be lower in dense star-forming regions like the Trapezium
cluster than in sparser regions like Taurus, the binary frequency in
the field constrains the relative numbers of stars that could have
originated in these different types of regions.  A simple recipe was
proposed here to account for the field population: take two parts of a
`Trapezium' population, add one part of a `Taurus' population, and the
mix reproduces quite well the binary statistics of the field.  The fact
that the field population resembles the Trapezium cluster more closely
than the Taurus region in its binary frequency suggests that most field
stars originated in clusters like the Trapezium cluster.

   Infrared observations of the newly formed stars embedded in molecular
clouds show that most stars do indeed form in clusters.  In the Orion
clouds, for example, most of the newly formed stars are located in
several embedded or partly obscured clusters, of which the Trapezium
cluster is just the largest.  The IMFs inferred from the infrared
observations of these young clusters are in good agreement with the
field IMF, again consistent with the possibility that most field stars
are formed in such clusters.  Most of these embedded clusters must
however be very short-lived, and only the largest ones can survive for
any length of time as open clusters.  The Trapezium cluster might evolve
into something like the Pleiades cluster, probably losing most of
its stars in the process, but the smaller Orion clusters will soon
evaporate.  Even the Pleiades cluster will not survive for long compared
with the age of our Galaxy, so the oldest open clusters that we now see
must be just the surviving remnants of a once much larger population.
Also, we heard that in order to account for the properties of the oldest
open clusters, there must once have been many clusters much more massive
than those we see now, with masses of tens of thousands of solar masses.
The first clusters formed in the Galactic disk might therefore have had
masses more like those of globular clusters than those of present-day
open clusters.

   On larger scales, molecular clouds produce associations of young
stars that may contain several clusters or subgroups as well as a more
dispersed population that is already beginning to dissolve into the
field.  The classical OB~associations are now known to contain not only
massive stars but also many low-mass stars, and to be the birth sites
of most stars of all masses.  Many OB~associations contain subgroups
that were formed in several distinct episodes of star formation, and in
some cases the most recent star formation has occurred at the edge of
an association in gas that appears to have been compressed by expanding
shells produced by the earlier episodes of star formation.  For example,
star formation in both the $\rho$~Ophiuchus and Orion regions is now
occurring in filamentary clouds whose appearance suggests that they are
being compressed and ablated by outflows from the previous centers of
star formation in the region.  Star formation may thus sometimes trigger
further star formation in nearby gas, but the star-forming clouds are
soon also destroyed by these same effects, and star formation is then
shut off.  Thus the feedback effects of star formation are complicated,
and it may not always be clear whether the net effect is positive or
negative.  Generally, star formation may occur wherever large-scale
gas motions cause the interstellar gas to pile up into dense molecular
clouds, but the gas motions involved are often turbulent and chaotic,
and it may then not be possible to identify a unique cause or
triggering effect for each episode of star formation.

\section{Moving Streams as Remnants}

   When associations disperse, they retain some kinematic coherence
because their internal motions are relatively small, and they then
become moving groups or streams.  The weak-lined T~Tauri stars found
by X-ray observations that were discussed at this meeting probably
represent some of the remnants of associations that are just now
dissolving into the field and becoming moving groups.  Several young
moving groups have been known for some time that are associated with
clusters such as the alpha Perseus and Pleiades clusters, and a few
older moving groups have also been identified.  Most if not all of the
field population probably consists of the remnants of old associations
that have become moving groups, but the identification of old moving
groups is difficult because they eventually form bands and not just
clumps in velocity space.  Therefore very complete and accurate data are
needed to sort them out, but much progress should become possible with
planned future instruments capable of gathering such data.

   It has also become clear in recent years that the Galactic halo
contains moving streams that are the debris of disrupted halo systems
or satellite galaxies.  A large fraction of the halo could belong to
a relatively small number of streams created by the disruption of a
relatively small number of satellites.  Some of the nearby dwarf
spheroidal galaxies have extended envelopes that are being tidally
dispersed, and the debris from these systems could itself account for
a significant fraction of the halo.  It is also possible that most of
the halo field stars could have originated in many small and relatively short-lived clusters.  The recently discovered Sagittarius dwarf galaxy,
which is just now being disrupted by interaction with our Galaxy,
provides a clear `smoking gun' example of a satellite that is being
dispersed into an extended moving stream and thus is adding both stars
and clusters to the Galactic halo.  The dwarf spheroidal galaxies
presently observed around our Galaxy are probably just the fading
remnants of a once much more prominent system of satellites, most of
which have by now merged with our Galaxy to build up the halo and
perhaps also the bulge and thick disk components.

\section{The Galactic History of Star Formation}

   There has been much interest in reconstructing the history of star
formation in our Galaxy and others from their stellar age distributions,
and new results were presented here for the star formation history of
our Galaxy, the Magellanic Clouds, and several other Local Group dwarfs.
The non-uniform age distribution of the stars in the Galactic disk
suggests that a number of episodes of enhanced star formation activity
occurred at intervals of several Gyr in the past, the most recent one
having occurred a few Gyr ago.  It is intriguing that both the Large and
the Small Magellanic Clouds also show evidence for large increases in
their star formation rates at about the same time a few Gyr ago, and it
has been suggested that all of these events might have been caused by a
close encounter among the three galaxies a few Gyr ago.  Some anomalies
in the spatial distribution of gas and young stars in our Galaxy, such
as the Gould Belt and other corrugations or warps, might also have been
caused by recent interactions with small companions or infalling
gas clouds, but the origin of these features is not yet well understood.

   Additional information about the history of star formation in our
Galaxy is provided by the chemical abundances of stars as a function
of age.  The recent results presented here confirm that, although the
average metallicity of the stars in the Galactic disk has increased
smoothly and monotonically with time, there is a substantial scatter in
the age-metallicity relation that probably implies significant chemical
inhomogeneities in the interstellar medium.  The classical `G-dwarf
problem', i.e.\ a paucity of metal-poor stars relative to the
predictions of simple models, also persists in the most recent data.
Gas infall, interactions with companions, starburst events, and
variability of the IMF could all play some role in explaining these
observations, but galactic chemical evolution remains a poorly
understood subject in all but the broadest outlines; simple models
clearly fail to account for all of the observations, but the data are
not yet sufficient to constrain adequately the many more complex models
that have been proposed.

\section{Starbursts and Superclusters}

   Elsewhere in the universe, there is abundant evidence for bursts
of star formation associated with interactions between galaxies, and
the most extreme starbursts appear invariably to be caused by violent
interactions or mergers.  These interactions redistribute the gas in
galaxies, typically piling it up near the center where an intense
burst of star formation results.  Spectacular central starbursts are
especially favored if the colliding galaxies have prominent bulges,
while in less centrally concentrated systems the resulting star
formation activity may be more widespread.  The frequency of violent
interactions and starbursts increases strongly with redshift, suggesting
that the earliest stages of galaxy formation and evolution might have
been dominated by starbursts triggered by interactions.  By contrast,
star formation in most present-day galaxies is a relatively quiescent
and orderly process, and interactions and triggering events now play a
less important role.  Present-day star formation probably results mostly
from small-scale gas motions that are more random in nature and are not
the result of any obvious triggering effect, although local violent
events such as the formation of expanding shells by multiple supernova
explosions may sometimes play a significant role in triggering local
star formation.

   There was much interest at this meeting in the formation of very
luminous star clusters or `superclusters'.  These objects are found in
regions of exceptionally vigorous star formation such as bright spiral
arms, starburst regions, and galactic nuclear regions where star
formation is highly concentrated.  Active star formation seems generally
to be highly clustered, and the fraction of stars that form in luminous
clusters increases with the local star formation density in galaxies.
The most luminous young clusters are of particular interest because of
the possibility that they may represent young globular clusters; if
they do, this would mean that the formation of globular clusters is
continuing in some locations and thus is open to observational study.
Some luminous young clusters are found to have the sizes and masses
of globular clusters, so they could indeed be young globular clusters.
However, it is not known whether they will actually survive for a Hubble
time; this depends sensitively on their IMF, since survival for a long
time is possible only if a cluster is sufficiently dominated by low-mass
stars.  If a cluster contains too many massive stars, it will lose too
much mass as these stars evolve, and as a result it will soon evaporate.
Even the standard IMF discussed above contains barely enough low-mass
stars for a cluster to survive for a Hubble time, so it could well be
that most of the luminous young clusters do not survive for a Hubble
time.  In this case the observed globular clusters could be, like the
oldest open clusters, just the disappearing remnants of a once much
larger population.

\section{Star Formation in the Smallest Galaxies}

   Low-mass disk and irregular galaxies tend to be relatively isolated,
and they lead very quiet lives, forming stars at a leisurely pace and
retaining a large amount of gas.  These systems may have settled into
a state of marginal stability, and they may form their stars mostly
in small local events wherever the self-gravity of the gas becomes
sufficiently important.  However, they may also be particularly
vulnerable even to relatively mild interactions, which apparently can
trigger galaxy-wide starbursts in them.  Some dwarf galaxies such as
the `blue compact dwarfs' seem to be able to produce central starbursts
without any obvious external cause, possibly as a result of a slow
accumulation of gas at the center until a stability threshold is
surpassed.  Starbursting dwarfs of both types may account for some
of the apparent excess of blue galaxies that is seen at intermediate
redshifts; if so, a general conclusion might be that giant galaxies
form most of their stars at large redshifts, while dwarf galaxies form
most of their stars at more modest redshifts.

   The Local Group dwarfs have been much studied with respect to their
star formation histories, which turn out to be very diverse.  The
gas-poor dwarf spheroidal galaxies around the Milky Way have no young
stars, yet most of them have significant intermediate-age populations
indicating that they continued to form stars for many Gyr after their
formation.  The fraction of intermediate-age stars tends to increase
with distance from the Milky Way, and the Fornax dwarf, one of the most
distant ones, contains some stars younger than 1~Gyr in age.  The two
largest dwarf spheroidals, Sagittarius and Fornax, also have their own
globular cluster systems.  The star formation rates in these systems
have generally decreased with time, but in some cases there have
been large fluctuations or even recurring episodes of star formation
separated by inactive periods.  These diverse behaviors may reflect the
fact that small systems are vulnerable to a number of effects that can
either suppress or enhance star formation; supernova-driven gas loss
and sweeping by motion through a hot ambient medium may remove gas
and suppress or shut off star formation, while interactions with other
systems or infall of new material may reinvigorate star formation for
a brief time.  But in general, rather than telling us about what drives
star formation, these systems may be telling us more about how star
formation sputters out when galaxies run out of gas.

\section{The Fading Fireworks}

   A general theme of much of this meeting has been that star formation
occurs mostly in discrete events or bursts that produce stellar
aggregates of all sizes ranging from small groups to galaxies.  These
events can have many causes, but in all cases they involve the rapid
accumulation of gas in a small volume of space where self-gravity
eventually overwhelms all other effects and causes rapid collapse to
occur.  The resulting bursts of star formation are brief but spectacular
events that can produce brilliant displays of massive stars and luminous
clusters.  The remaining gas is then quickly dispersed, and soon the
young stars and clusters are themselves mostly dispersed, creating the
moving streams of which galaxies are built.  Much of what astronomers
study can be understood as the fossil remnants of past starburst events
which were once more common than they are now.  We presently see mainly
the fading embers of the cosmic fireworks show, but it has left us a
rich fossil record to study, and we can compare this record with the
star formation activity that we can still observe in many locations.
Increasingly, we can also use high-redshift observations to study
directly the star formation activity that occurred at earlier stages in
the evolution of galaxies, and thus check the inferences drawn from the
fossil record.  The evidence that is beginning to accumulate suggests
that much of what happened at earlier times can be understood as
scaled-up versions of the present-day star formation processes that
were the main topic of this meeting.

\end{document}